# Hankel Spectrum Analysis: A novel signal decomposition method and its geophysical applications

Kunpeng Shi[1], Hao Ding[1,2*]

[1]*Department of Geophysics, School of Geodesy and Geomatics, Key Laboratory of Geospace Environment and Geodesy of the Ministry of Education, Wuhan University, 430079, Wuhan, China*

[2]*Hubei Luojia Laboratory, Wuhan 430079, China*

**Corresponding address**: dhaosgg@sgg.whu.edu.cn

Key points:

1. Hankel spectrum analysis (HSA) is proposed for more exactly decomposing and parameterizing non-stationary geophysical signals

2. Three phase jumps of the Chandler wobble accompanied by sudden drops in instantaneous amplitude and period are detected for the first time

3. We report the Earth's oblateness $\Delta J_2$ of a ~18.6yr tidal signal, an ~10.5yr possible solar cycle, and multiple global gravity anomaly peaks.

## Abstract

To analyze non-stationary harmonic signals typically contained in geophysical observables is a quest that has seen continual advances in numerical techniques over the decades. In this paper, based on transient $z$-pole estimation (in Hankel matrices), a novel state-space analysis referred to as Hankel

Spectral Analysis (HSA), was developed. Depended on the Hankel total least square (HTLS), the HSA incorporates truncated singular value decomposition (TSVD) and its shift-invariant property in robustly decomposing the closely-spaced sinusoids. Resorted to a sliding window processing, HSA can be used to analyze non-stationary sequential structures, in the support of consecutive quaternary parameters $\{A_i, \alpha_i, f_i, \theta_i\}$. Based on a series of experiments with special features commonly in real measurements, the availabilities of HSA in complex harmonic constituents (e.g., the time-variant amplitude/frequency, mutation, the episodic recording signals) with low Signal-to-Noise Ratio are confirmed. In real applications, we use HSA to analyze both global geophysical observables, including polar motion (PM) and earth's dynamic oblateness ($\Delta J_2$), and some new findings are obtained. In the PM series since the 1900s, a total of triple jumps from Chandler wobble (CW) are firstly confirmed; and all of them are synchronized by the sharp decrease of Chandler intensity and period. In the $\Delta J_2$ series, two decadal signals (18.6 yr, 10.5 yr) are identified to be associated with the tide effect, and solar activity; and its interannual-to-decadal oscillations contribute to multiple global gravity anomalies. These findings implied the great potential of the HSA in searching hitherto signals of geophysical observations.

# 1 Introduction

In geophysics and geodesy, different types of stations distributed globally have accumulated a large number of continuous records. The ability to detect the periodic and quasi-periodic signals of avail is quite essential because they contain rich information and underlying mechanisms. Harmonic analysis, nevertheless, is challenging for some limitations. First, conventional tools' poor sensitivity and spectral resolution restricted the hybrid signals detected and separated, especially in the close-frequency bands (Ding et al., 2015a, 2018). Second, investigating covert long-period signals and giving geophysical interpretation remains indeterminacy by the background noise inherent in geodetic time series (Chao et al., 2020). Besides, evaluation of the structural anomaly is another complicated task, which is diagnosed by consecutive spectrum analysis (Malkin and Miller, 2010). Up to now, defining an adaptive analysis method that is original from the time series itself and then satisfies the applications in Geo-field is still a burgeoning topic (Wdowinski et al., 1997; Hennenfent et al., 2006).

The harmonic analytic techniques are mainly categorized as signal recognition and restoration process. For the former, various numerical methodologies have been devised, many of which have roots in the conventional Fourier harmonic analysis (Slepian, 1978) and some others are non-Fourier-based methods (e.g. Lomb-Scargle periodogram, product spectrum, cross-spectrum) (Thomson 1982; Carlin & Louis 2011). However, some disadvantages limit their applicability in practice; for instance, the spectral analysis methods are not accurate enough for identifying harmonics in damped cosines\sines, and a longer length and adequate sampling interval are also necessary to obtain enough frequency resolution. Aimed at accurate signal identifications, the modern geodetic field has appealed for higher sensitivity and resolution than ever. Lately, some spectral analysis methods considering sequential structure (like MEM and AR-$z$ method (Ding & Chao, 2018)) arose numerous researchers' interest. These methods implied broad prospects of harmonic analysis if we relied on the special structure of the time series itself.

Another vital work to process geophysical observations is signal restoration, including signal decomposition or reconstruction. Among many numerical restorations, the famous empirical mode decomposition (EMD; Huang et al. (1998)) and its improved version (such as Ensemble EMD (EEMD), Wu and Huang, 2009) obtained many applications to decompose signals into symmetric, narrow-band waveforms named intrinsic mode functions (IMFs). Though improved the time-frequency resolution, they still inevitably suffer from the undesirable shortcomings commonly in numerical decomposition, including but not limited to tail effect, frequency aliasing, and mode mixing (Tang et al., 2012). How to overcome these defects has gradually become a critical problem (Jiang and You, 2022). Before the EMD-based methods, another well-known algorithm of Singular Spectrum Analysis (SSA) (Vautard et al., 1992) is developed. In geodetic and geophysics, the SSA method has been widely applied in extracting the principal components; whereas the singular values coinciding have to be considered. In this situation, the empirical statistics in determining eigenvectors may make disjoint sets of singular values and lead components mixed (Golyandina and Shlemov, 2013). Another issue that cropped up was that the SSA is completely data-driven, unfollowing any predefined mathematical model so that the reconstructed signals generally cannot be used to infer the underlying physical mechanism. This problem is commonly occurred in the non-parametric analysis method (Wang et al., 2007), causing the decomposition failure both using SSA or EEMD in short-length observations to recover the long-period signals.

To enhance the harmonic analytic ability for non-stationary signals, in this study, we propose a new signal decomposition method, referred to as the Hankel Spectral Analysis (HSA). The core procedure of the HSA is using the Hankel Total Least Squares (HTLS) to solve the z-poles exactly. It's Prony (1795) who firstly estimated z-poles in the autoregressive (AR) model to obtain the parameters of a sum of $K$ exponentially damped sinusoids. However, such a model is sensitive to the initial condition and more than one pole usually lies outside the unit circle on the complex z-plane in noisy data. This means that the amplitude of the corresponding reconstructed components will abnormally exponentially be amplified with time (Chao, 1993). Since then, other z-poles estimation algorithms have been successively developed, yet still become incapable in some sinusoids with closely spaced, low signal-to-noise-ration (*SNR*) or a small number of samples (Hua and Sarkar, 1990; Kumaresan and Tufts, 2003). Chen and Huffel (1996) have explained that the Hankel *z*-poles solution accuracy can be improved upon using the HTLS technique. Firstly, the shift-invariance property is transformed from a Vandermonde vector basis to the SVD-Hankel dominant subspace, and a system of equations will be built afterward. Secondly, the TLS solution of this system is computed to yield the frequencies and damping factors (Huffel, 1998). Finally, a set linear parameter including amplitude, frequency, phase, and damped factor will be obtained (Chen et al., 1997). In the subspace-based processing framework, the HTLS is a TLS variant of Kung et al. (1983)'s algorithm. In this study, we will use the HTLS to construct the HAS method. Resorted to the sliding window, its applications are popularized in retrieving transient structure to non-stationary harmonics, making the above troubles of existing numerical decomposition settled possible.

In the following, to demonstrate the superiority of the HSA, we shall conduct 5 synthetic experiment cases for the identifications and decompositions of harmonic signals with intense time-dependent, low-*SNR*, discontinuous, or mutated structures to demonstrate; as comparisons, we will simultaneously use the other two commonly used nonlinear decomposition methods, the EEMD and SSA. After then, we shall apply the HSA to investigate the global geophysical observables of the Earth, including the Chandler wobbles jumps in Polar motion (PM), and long-period variabilities in the Earth's oblateness ($\Delta J_2$). The obtained findings include weak signal detection, principal component extraction, and low-frequency identification. Some sequential characteristics that are unresolved hitherto will be exhibited, which will reveal fine structures in the underlying physics.

## 2 Methodology

The real-world observations can be modeled as a sum of exponentially damped sinusoids (Dahlen & Tromp, 1998), which can be approximated by the HSA technique. In this section, the Hankel Spectrum Analysis from the principal solution of Hankel-$z$ poles, to the sliding window process and related spectrum will be detailed description in succession.

### 2.1 Hankel Total Least Squares

Assuming a single, noise-free observation containing $N$ complex samples $x(n)$, $n= 1,2, ..., N$. Following previous studies, the time series can be modeled as a finite sum of $K$ different exponentially damped complex sinusoids:

$$x(n) = \sum_{k=1}^{K} A_k e^{j\theta_k} \cdot e^{(\alpha_k + j2\pi f_k)T_s(n-1)} = \sum_{k=1}^{K} c_k \cdot z_k^{n-1}. \tag{1}$$

Approximation of signal $x(n)$ can be expressed by $K$ different components, in which $A_k$ is the initial amplitude in the same units as $x_n$, $\alpha_k$ is the damping factor, $f_k$ is the frequency, $T_S$ is the sampling period of signal $x(n)$ and $\theta_k$ is the initial phase in radians. Therefore, $x(n)$ is characterized by the parameters $A_k$, $\alpha_k$, $f_k$ and $\theta_k$ ($k =1, …, K$). $c_k$ is the time-independent component (residues) and $z_k$ is the time-dependent component (poles). The HTLS analysis has a good performance in time consumption and noise suppression. The procedures are illustrated in Fig. 1 and we will now explain how this model can be handled in a subspace framework.

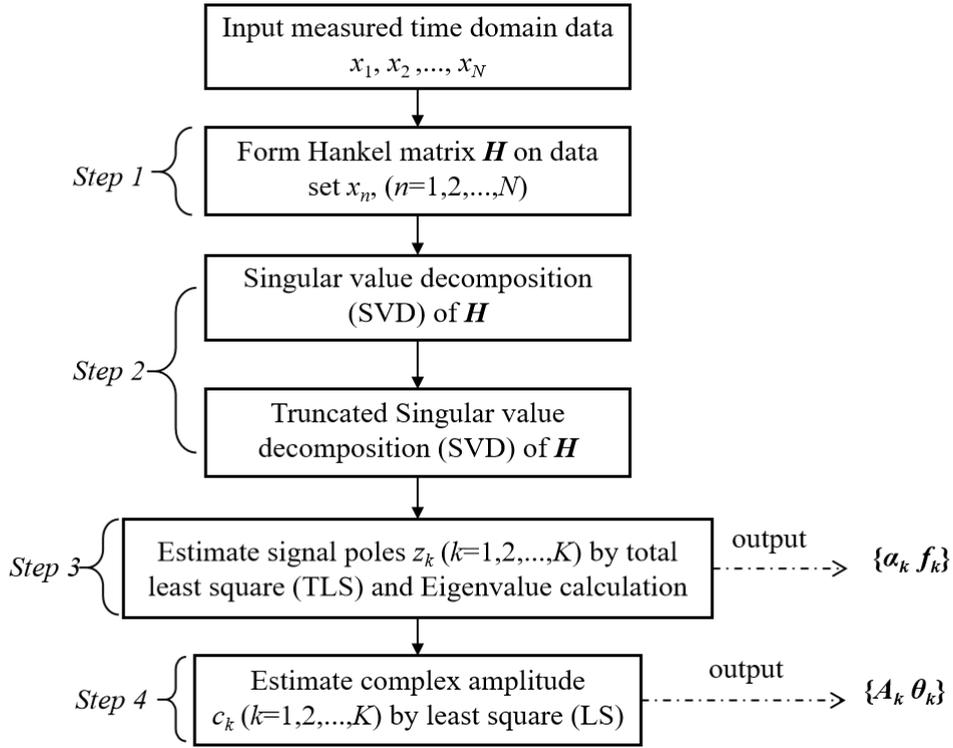

**Figure 1.** Parameter resolution scheme by the Hankel TLS

**Step 1**. *The construction of the Hankel matrix*

In the utility of the collected $N$ sampling data, the $L \times P$ Hankel matrix is arranged as

$$H = \begin{bmatrix} x_1 & x_2 & \cdots & x_P \\ x_2 & x_3 & \cdots & x_{P+1} \\ \vdots & \vdots & \ddots & \vdots \\ x_L & x_{L+1} & \cdots & x_N \end{bmatrix}_{L \times P}. \quad (2)$$

where the $P$, equivalent to the $p$ order in Prony analysis (Johnson et al., 2000) is the determinant for the Hankel matrix. The pair of $(L, P)$ is chosen such that $N = L + P - 1$ with $P \gg K$, and the $H$ matrix can be unfolded by a product form as:

$$H = \begin{bmatrix} 1 & 1 & \cdots & 1 \\ z_1^1 & z_2^1 & \cdots & z_K^1 \\ \vdots & \vdots & \ddots & \vdots \\ z_1^{L-1} & z_2^{L-1} & \cdots & z_K^{L-1} \end{bmatrix} \begin{bmatrix} c_1 & & & \\ & \ddots & & \\ & & \ddots & \\ & & & c_K \end{bmatrix} \begin{bmatrix} 1 & z_1^1 & \cdots & z_1^{P-1} \\ 1 & z_2^1 & \cdots & z_2^{P-1} \\ \vdots & \vdots & \ddots & \vdots \\ 1 & z_K^1 & \cdots & z_K^{P-1} \end{bmatrix}. \quad (3)$$

Eq. (3) can be rewritten as $H = SCT^{\mathrm{T}}$. This is called a Vandermonde decomposition in which the poles $z_k$ are also called the *generators* and the superscript T denotes the transpose. The matrices $S$ and $T$ possess a shift-invariance property that can be expressed as:

$$\begin{aligned} S_\downarrow Z &= S^\uparrow \\ T_\downarrow Z &= T^\uparrow \end{aligned} \quad (4)$$

where the ↑ (↓) arrow stands for deleting the top (bottom) row of the considered matrix, and $Z=\mathrm{diag}(z_1, z_2, \ldots, z_K)$.

**Step 2**. *The Truncated Singular value decomposition*

An SVD-decomposition is employed to solve the rank deficiency of $H$:

$$H = U\Sigma V^{\mathrm{H}} = \begin{bmatrix} \hat{U} & U_0 \end{bmatrix} \begin{bmatrix} \hat{\Sigma} & \\ & \Sigma_0 \end{bmatrix} \begin{bmatrix} \hat{V} \\ V_0 \end{bmatrix}^{\mathrm{H}}. \quad (5)$$

where ($U_{L\times L}$ $V_{P\times P}$) is called left and right singular orthogonal matrices pair corresponding to the diagonal matrix $\Sigma$ which is consisted of singular values in decreasing order $\{\sigma_1, \sigma_2, \ldots, \sigma_h\}_{h=\min(L, P)}$. The superscript H denotes the Hermitian transpose. In the case of free noise $\Sigma_0$ is a null matrix, but once the signal is corrupted by noise, a full matrix of $\Sigma_0$ has occurred. Here a truncated SVD of $H$ will be invoked as: $H = \hat{U}_{L\times K} \hat{\Sigma}_{K\times K} \hat{V}^{\mathrm{H}}_{K\times P}$ in the presentation of the best $K$-approximation. In practice, the $K$ order is related to the number of harmonics and usually can be obtained by inflection or collapse of a singular spectrum.

**Step 3**. *The total least squares for $z_i$ poles*

In the noise-free case, $\hat{U}$ equals $S$ up to a multiplication by a square non-singular matrix $Q_{K\times K}$:

$$\hat{U} = SQ. \quad (6)$$

Correspondingly, the multiplication can be further expanded in the form as:

$$\hat{U}^\uparrow = S^\uparrow Q$$
$$\hat{U}_\downarrow = S_\downarrow Q \qquad (7)$$

where the matrix pair $(\hat{U}^\uparrow \ \hat{U}_\downarrow)$ is similar to that of $(S^\uparrow \ S_\downarrow)$. In the shift-invariance property, an meaningful expression is produced when combined Eqs. (4) and (7) as:

$$\hat{U}^\uparrow = \hat{U}_\downarrow Q^{-1} Z Q = \hat{U}_\downarrow \tilde{Z}. \qquad (8)$$

In Eq. (8), $Z$ and $\tilde{Z}$ share the same eigenvalue $\{z_1, z_2, \ldots, z_K\}$. The TLS solution of the overdetermined set of linear equations is used as:

$$Z = -W_{12} W_{22}^{-1}. \qquad (9)$$

where the submatrix pair $(W_{12} \ W_{12})$ can be acquired when the SVD-decomposition for the matrix $[\hat{U}_\downarrow \ \hat{U}^\uparrow]$ is performed as

$$[\hat{U}_\downarrow \ \hat{U}^\uparrow] \stackrel{SVD}{=} Y \Gamma W^H. \qquad (10)$$

where

$$W = \begin{bmatrix} W_{11} & W_{12} \\ W_{21} & W_{22} \end{bmatrix}. \qquad (11)$$

The behavior of double SVD-decomposition in the TLS solution can effectively minimize the correction and make the dataset compatible. After the roots $z_k$ be computed, then the damping factor ($\alpha_k$) and frequency ($f_k$) can be obtained as:

$$\begin{cases} \alpha_k = \dfrac{\ln|z_k|}{T_s} \\[2ex] f_k = \dfrac{\tan^{-1}\left[\dfrac{\mathrm{Im}(z_k)}{\mathrm{Re}(z_k)}\right]}{2\pi T_s} \end{cases} \qquad (12)$$

**Step 4**. *The least-squares solution for $c_i$ residues*

Solve the original set of linear equations to yield the exponential amplitude and sinusoidal phase estimates. First, the initial exponential model ($\mathbf{Z}_{N \times K} \cdot \mathbf{c}_{k \times 1} = \mathbf{x}_{N \times 1}$) is solved:

$$\begin{pmatrix} z_1^0 & z_2^0 & \cdots & z_K^0 \\ z_1^1 & z_2^1 & \cdots & z_K^1 \\ \vdots & \vdots & \ddots & \vdots \\ z_1^{N-1} & z_2^{N-1} & \cdots & z_K^{N-1} \end{pmatrix} \begin{pmatrix} c_1 \\ c_2 \\ \vdots \\ c_K \end{pmatrix} = \begin{pmatrix} x_1 \\ x_2 \\ \vdots \\ x_N \end{pmatrix}. \quad (13)$$

In common practical cases $N > K$ and, in this situation, the systems are overdetermined (more equations than unknowns). The linear system can be approximated using the LS or TLS methods (Horg and Johnson, 2012; Rodriguez et al., 2018). Finally, the $c_k$ values yield the amplitude and phase $\{A_k\ \theta_k\}$:

$$\begin{cases} A_k = |c_k| \\ \theta_k = \tan^{-1}\left[\dfrac{\operatorname{Im}(c_k)}{\operatorname{Re}(c_k)}\right] \end{cases}. \quad (14)$$

**2.2 The sliding window process**

In the real geoscience applications, the studied signals are usually time-varying and even undergo sudden variations due to the occurrence of excitation events. A set of parameters described above cannot deal with such situations, so tracking harmonic properties become impossible. To overcome these drawbacks, the HTLS in a shift window fashion is proposed instead of a linear process.

The time series is divided into short overlapped time windows, and each one is parametric analyzed (see Fig. 2). The TLS solution of the Hankel matrix is well suitable for the proposed technique since it was originally developed to work with a small number of data samples. Different from some windowed methods (e.g., the Short-time Fourier, wavelet basis) that a design trade-off must be made between the time and frequency resolution (Rezaiesarlak et al., 2013), the HSA provides very high-frequency resolution even for very short time windows. It can considerably reduce the computational cost together with time acquisition. When the specific window shifts back (step length=1), we can track the attributes of non-stationary harmonic with high accuracy.

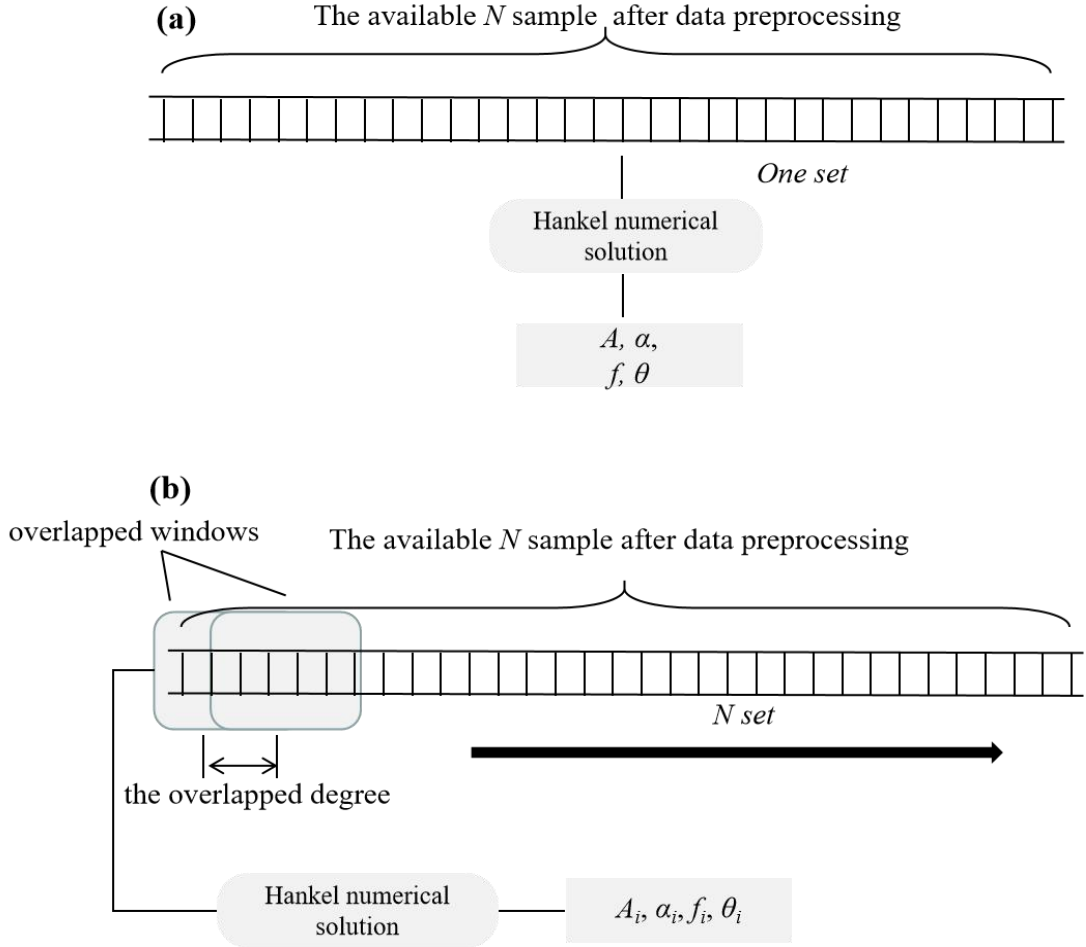

**Figure 2**. Description of the linear HSA (a); and its sliding process(b).

**2.3 HSA spectrum**

Using an analogy to the discrete Prony spectrum (Mitrofanov and Priimenko, 2015), we can call this set "the discrete HSA spectrum" $\Re_T(K)$, i.e.,

$$\Re_T(K) = \{f_k, \theta_k, A_k, \alpha_k\}_{k=1}^{k=K}. \tag{15}$$

For a window width defined by $T$, the HSA decomposition can be carried out. Here its time interval is characterized by $Ts$. The parameter $T$ plays an important role in the HSA spectrum, and the $\Re_T(K)$ depends on the order of approximation, $K$. When using the HSA decomposition, the near-orthogonality of damped sine/cosine functions cannot be neglected. This property allows the signals to be decomposed into merely several components corresponding to narrow or even variant frequencies. The instantaneous transition of recorded data at each point, to the consecutive HSA spectrum, can be represented as follows:

$$observed\ series \xrightarrow[\text{sampling}]{\text{data}} \{x[t]\}_{t=1}^{t=N} \xrightarrow[\text{by } i,T]{\text{Selection window}} \{x[n]\}_{n=i}^{n=i+T} \xrightarrow[\text{transform}]{\text{HTLS}} \Re_T(K_i)\ . \quad (16)$$

where $\{x[t]\}_{t=1}^{t=N}$ is the total set of discrete data. For the specific window width $T$ (set in advance), the time interval is determined for the HTLS numerical solution. As the position $i$ varies covering the entire series ($1 \leq i \leq N-T$), the parametric transform in the part of $\{x[t]\}_{t=1}^{t=N}$ is continuously performed till the "consecutive HSA spectrum" $\Re_T(K_i)$ is obtained.

## 2.4 The model parameters

According to Hauer et al. (1990), the length of the sliding window $T$ should be at least one and half times of the lowest frequency of the target frequency band. For superimposed signals in close bands, the wider window and higher sampling frequency are necessary for enhancing the solvability of the Hankel matrix besides the smoothing of components.

The combination order of ($P$, $K$) is another important parameter pair for HSA analysis to define the Hankel-matrix $(N-P+1) \times P$ and harmonics numbers ($K/2$). A common principle is that the $P$ order should be no more than one-third of the selected data length (Johnson et al., 2000), no matter for the linear or shift window HSA. The ordered pair ($P$, $K$) could be increased together to accommodate much more modes. In our case studies, the initial $P$ order is selected by ~1/3 shift window length and *K* order is defined by the singular value collapse (*K*=4 for the sum of two sine-waves, for instance). It is challenging to make the first selection since the exact number of modes of a real system is hard to determine. If the pair (*P*, *K*) is found to be not high enough, they should be increased appropriately in the utility of the extended window length when necessary.

## 3 Experimental verification

In the presentation from globally to locally feature, the fine structure of time series has been explored. Many windowing algorithms can decompose the interesting series into different levels on multiple time scales. Unfortunately, with the basic function and decomposition scale set in advance, this method's efficiency largely depends on a predefined dictionary (Tary et al., 2014). In this study, the signal with the smoothness of amplitudes and frequencies is controlled in decomposing but uses the HTLS basis instead. In this section, five synthetic cases (cases 1-5) in multi-types structures consisting of time-varying of frequency and amplitude signals, low-*SNR*, episodic, mutation signals and the geophysical example of Polar motion simulation will be used to confirm the HSA effect with parameter spectrum.

## # Case 1: Amplitude-dependent signals

In this case, we use a simulated noise-free time series composed of time-dependent amplitude signals. Referring to Hou and Shi (2013), the synthetic series is written as:

$$x_i(t)=A_i(t) \cos(2\pi t \times f_i), x(t)=\sum x_i(t), i=1,2,3. \qquad (17)$$

where the $f_i$ (i=1,2,3) is the input signal frequencies of 0.05, 0.02 and 0.1 Hz, with the sampling interval of 1 s for 1000s length; and the time-varying $A_{1-3}(t)$ measures their corresponding cosine amplitude modulation (periods: 250, 500, 1000s). A comparison of HSA restored signals and input signals will be made to test its decomposition ability.

Fig. 3a is the waveform of this simulated time series, the three input variant-amplitude signals are shown in Figs. 3b-d (black curves), respectively. The ternary HSA parameter is set as wavenumber $K=6$ (3 pairs of harmonics), window length $T=100$ s and order $P=50$ to smooth reconstructed components. After using HSA in this simulated time series, the corresponding restored signals are shown in Figs. 3b-d (green curves), respectively. It can be found that the HSA perfectly decomposes the three input signals from the simulated time series. Such precise restoration is attributed to the invariance of spatial rotation of the HSA and the shift window used in it.

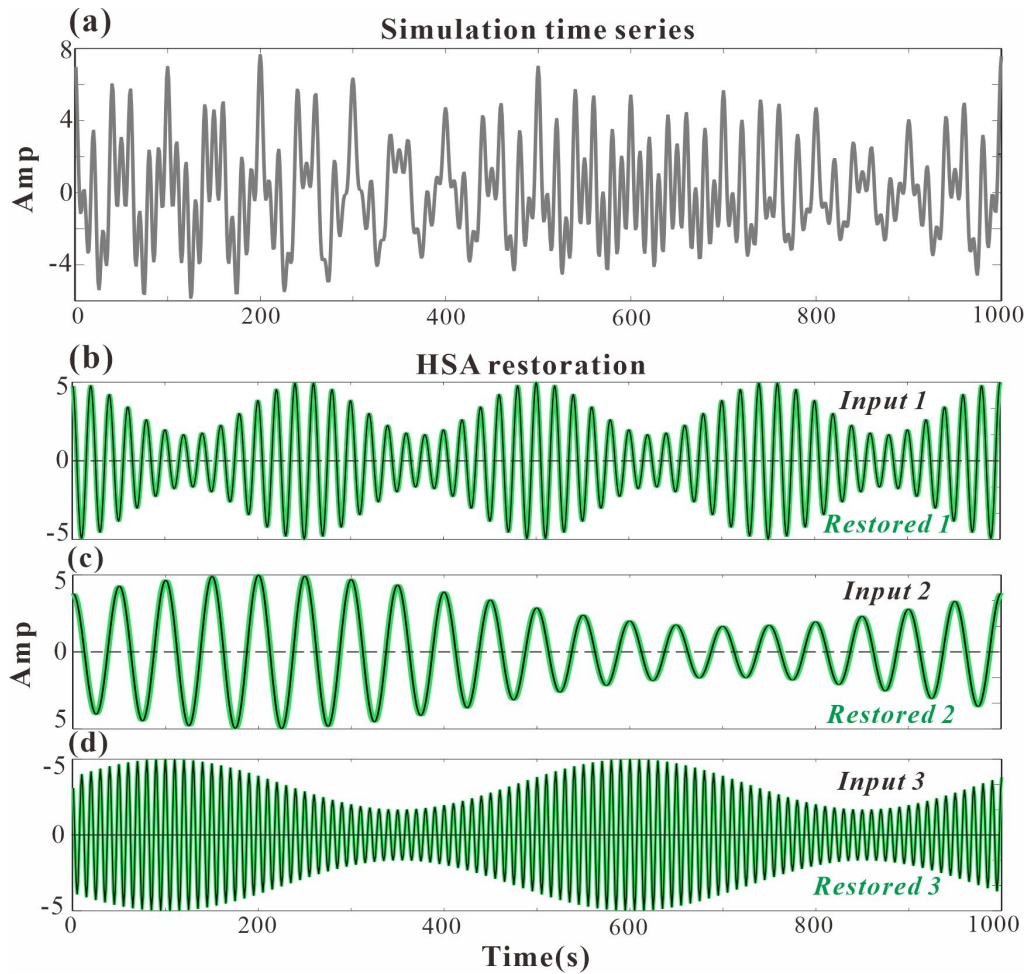

**Figure 3.** The simulated amplitude-dependent time series (a) and the comparison of input signals and HSA restoration (b)-(d)

# Case 2: Frequency-dependent signals

In case 1, the powerful ability of HSA to track the violent vibration in a variant-amplitude system has been revealed. In case 2, we will further simulate new time series, containing the linear (input signal 1) and sine frequency modulation (input signal 2), to verify HSA parameter recognition in the fast frequency variations. This simulation time series is superposed by $x=x_1(t)+x_2(t)$, and both input signals are expressed as:

$$x_1(t)=\sin(2\pi(30+20t)\cdot t). \tag{18}$$

$$x_2(t)=\sin(2\pi(5t+\sin(2\pi\times 0.3t))). \tag{19}$$

The sampling frequency is 500 Hz with a time interval of 0.002 s for 5 s, and the simulated time series are shown in Fig. 4a. According to the definition of the modulation system proposed by Ville et al. (1948), the instantaneous frequency of the input signal 1 ($f_1(t)$) and signal 2 ($f_2(t)$) can be derived easily as follows:

$$f_1(t) = \frac{1}{2\pi}\frac{d\theta(t)}{dt} = 5 + 40t. \tag{20}$$

$$f_2(t) = \frac{1}{2\pi}\frac{d\theta(t)}{dt} = 5 + 2\pi\times 0.3\times \cos(2\pi\times 0.3t). \tag{21}$$

where $\theta(t)$ is the phase function. As explained above, the instantaneous frequency of the signal also can be obtained by using HSA. The reconstructed signals and the corresponding derived parameter $\{f_i\}$ with time-varying features are shown in Figs. 4b-e, respectively. For this modulation system whose frequency varies with time, a recognition action is perfectly implemented by HSA tracking. It is clearly shown that the HSA can completely restore both input frequency-variant signals, and record their instantaneous frequencies. Other $z$-poles analytical techniques, e.g., the common Prony algorithm, have been previously integrated with wavelet base to acquire a comparative effect manifestation (Lobos et al., 2009). However, the difficulty of selecting a wavelet base and intense tail effect prevents its further popularization. With the high-accuracy identification of time-varying frequencies, the HSA approach with its simple, operable feature has a better application prospect.

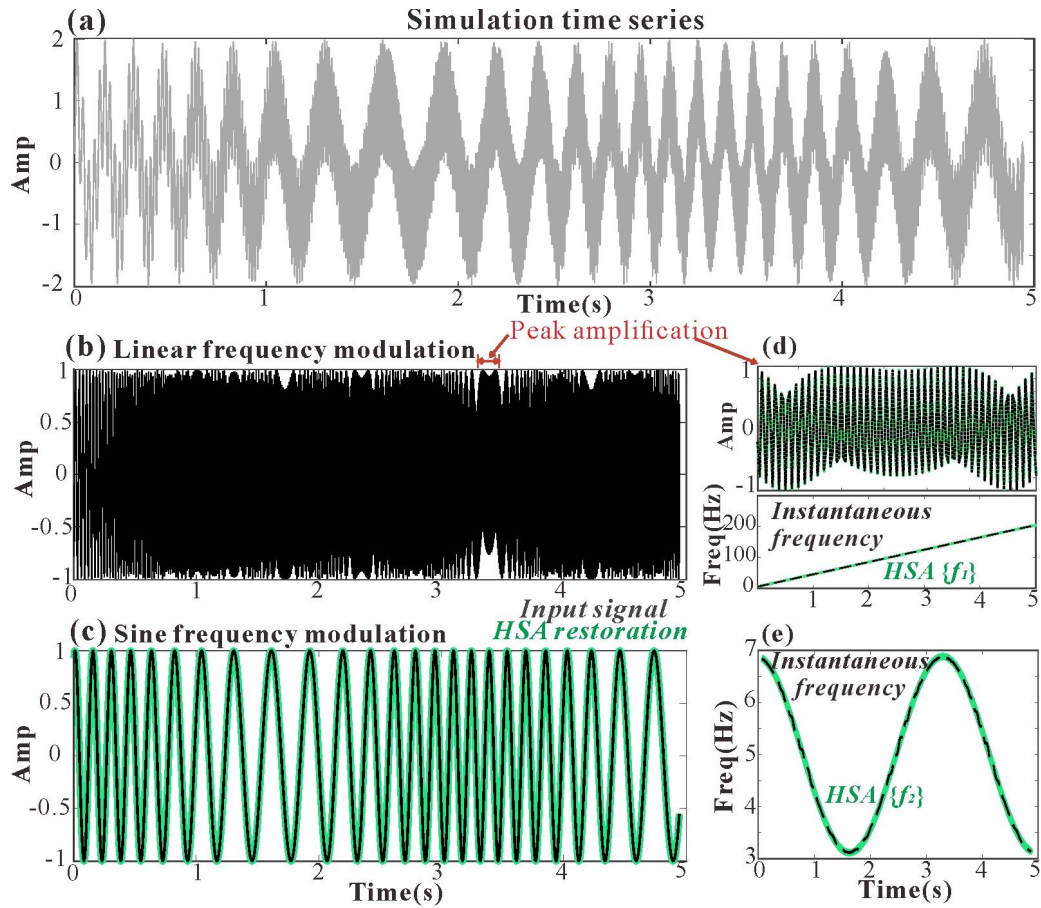

**Figure 4.** The simulated frequency-dependent time series (a) and the comparison of input signals and HSA restoration (b)-(c) with the local zooming and frequency identification (d)-(e).

# Case 3: Noisy time series

The double SVD-decomposition endows the HSA with the ability to minimize the time-varying signals correction even in high-intensity noise. Here we simulate a time series consisting of two different signals (see Fig. 5a, black curve), component 1 and component 2. The amplitude of component 1 changed from 0.084-0.115 with a fluctuating period near 365s (see Fig. 5b, black curve); while with an intense attenuation, the amplitude of component 2 declined from 0.146 to 0.046 and its variant periods are presented at 422 to 440s (see Fig. 5c, black curve). If there is no noise input into this simulated time series, after using HSA, the two input components can be restored with ultra-high precision (here we don't further show them). When the noise is input (the signal to noise ratio (SNR) is 5db), the simulated noisy time series is shown in Fig. 5a (gray curve). Fig. 5b shows the restored component 1 and its instantaneous periodic spectrum; Fig. 5c is similar to Fig. 5b but for Component 2. Those two figures demonstrated that HSA can accurately detect the parameters $\{A_i, \alpha_i, f_i, \theta_i\}$ of the two oscillation modes under the high-intensity noise. The waveform constructed by the HSA algorithm is highly equivalent to the components to be measured; and the HSA spectrum reveals the irregular periodic variations, signifying that the HSA can track transient structures of quasi-periodic signals even in a highly noisy environment.

Given that both input signals have time-varying amplitudes and periods, we also adopted well-used SSA and EEMD methods to decompose this simulated time series to verify the HSA further, but only a dissatisfied approximation was achieved (see the blue and red curves in Figs. 5b and 5c). Although we can identify two signals with ~365s and ~430s periods in the Fourier spectrum, no traditional method can almost completely isolate them from the time domain.

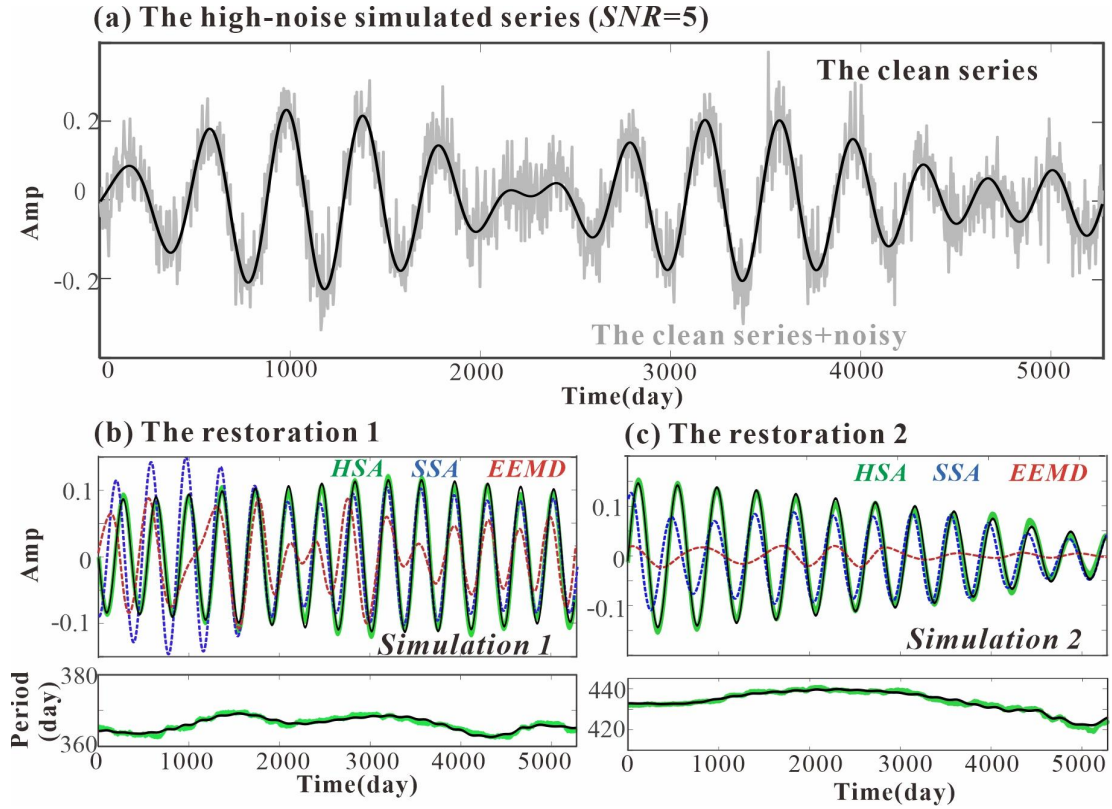

**Figure 5.** The HSA technology for simulated series polluted by Gaussian noise (a) to restore the component 1 (a) and 2 (b) using HSA, SSA and EEMD, with HSA periodic spectrum below.

# Case 4: Episodic signals

The episodic signals happening occasionally and not at regular intervals have widely existed in geo-field. e.g., episodic deformation in the volcano or tectonic process (Walwer et al., 2015). In that case, the creation of highly localized time-frequency decomposition is probably impractical in the traditional model (Wu et al., 2018). By contrast, a clear 'time-restoration' map even for signals in discontinuity is possible using HSA. In case 4, a simple simulated series will be produced to verify the HSA decomposition as:

$$x_i=\sin(2\pi t_i \times f_i),\ x=\sum x_i,\ i=1,2,3. \tag{22}$$

where $t_1$-$t_3$ is the time interval from 0-5 s, 0-1 s and 2-5 s, with the corresponding $f_{1-3}$ as 5 Hz, 10 Hz and 2 Hz, respectively. A sum of $x_1$ (continuous signal), $x_2$ and $x_3$ (episodic signals) are input for this simulation series.

Fig. 6 shows the simulated series and the HSA output of restored signals. Seeing Figs 6b-d, the spectrum analysis reveals that both slight *z*-poles solution abnormality occurred at junction points of components, whereas it does not affect the impressively identification and separation of discontinuous quasi- or periodic signals in an unstable system. Here, an overestimated state was allowed according to the initial exponential model (in the Vandermonde matrix form of Eq. 13). In the unknown wave numbers, a relatively large *K* value is recommended (e.g., *K*=10 in case 4) to recover the harmonics completely. For most local approaches in time-frequency analysis, the energy is not steadily distributed. Owning to the ability to track instantaneous parameters, the proposed HSA method can provide steady and compact reconstruction components, which sharpens the time domain distribution. Such time series are commonly observed in geoscience records; and in a rigorous mathematical frame, the HSA strategy can be used to facilitate recording episodic events of geophysical observations.

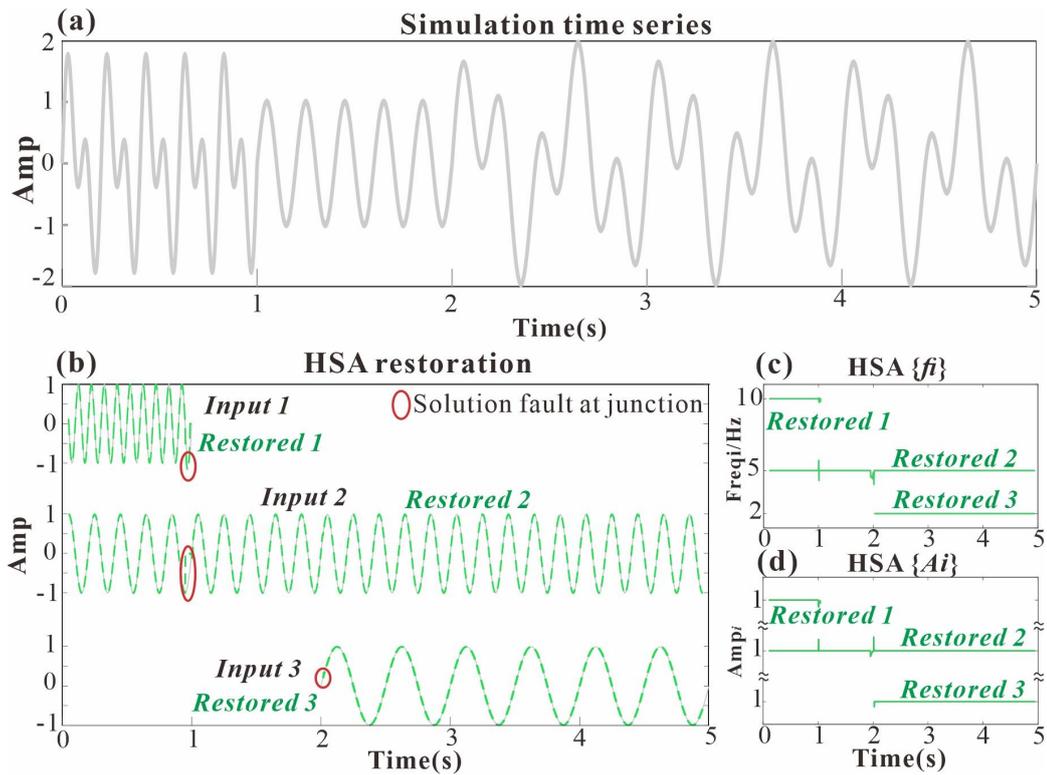

**Figure 6.** The simulated episodic time series (a) and the comparison of input signals and HSA restoration (b) with the frequency and amplitude identification (c)-(d).

# Case 5: Mutation signals

The additionally interesting HSA application, except for numerical decomposition mentioned above, is about observing the subtle structural mutation of time series in the parameter spectrum. We conduct the fifth experimental example (case 5) with mutation happened to demonstrate its tracking capabilities when a distorted signal emerged. The simulation series is sampled at the frequency of 500Hz for 5s length composed with baseband signal (20 Hz) and two other sidebands (24-22Hz, 16-18Hz) distorted at 2s. The proposed HSA was implemented to experiment with the composite signals with rather close frequencies.

Targeted at tracking the frequencies and amplitude of the searched components, the acquired series is divided into lots of short overlapped time windows (50 samples). In the process of gradual sliding backward, the quaternion parameters of each window are analyzed by the HSA method. Seeing the HSA spectrum in Fig. 7, the results depicted prove its high reliability in recording the distorted moment of weak sideband signals (red and blue lines) despite the baseband harmonic existing (black lines). The sideband frequencies in the simulation record are not constant; they present small jerks in frequency (2Hz at 2s) and large gaps in amplitude multiplication. In geodesy, the structural distortion in the measurements was always regarded as a huge challenge. This test confirms that mutation tracking can be performed exactly using HSA even with a low sampling rate.

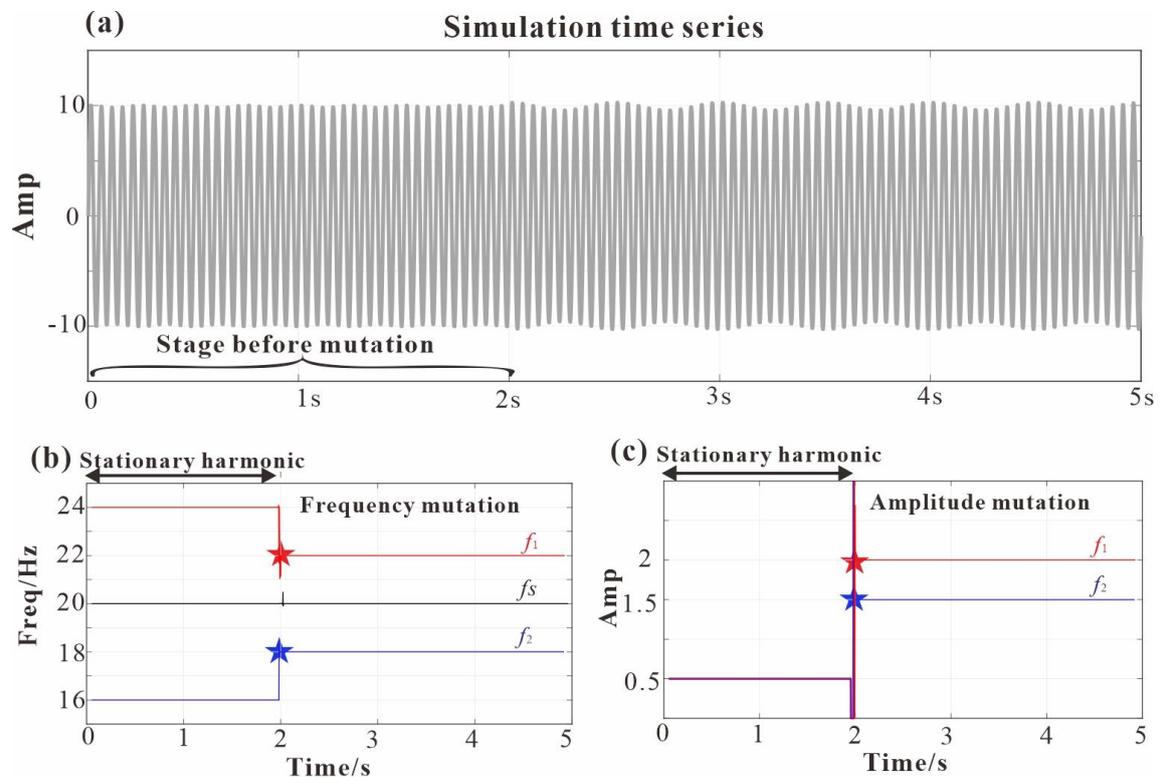

**Figure 7.** The frequency and amplitude tracking (sampling interval at 0.002 s for 5 s) for the composite signal occurred suddenly at 2s. The simulated time series are presented in (a) and HSA spectrum is shown in (b)-(c) for sidebands ($f_1$, $f_2$) and baseband signals ($f_s$), where the triangles represent structural mutations.

# 4 Geodetic applications

For the time-domain decomposition of an observed record, the traditional non-parameter approaches usually give an inappropriate description of the time variability of non-stationary signals. How to separate the discontinuous components precisely remains a pretty tough dispute in the geodetic field. In this section, applications for two typical geophysical time series, the Polar motion and dynamical oblateness $\Delta J_2$ will be carried out based on the HSA. For the Polar motion, we mainly focus on accurate stripping of the two main components, the Chandler wobble (CW) and the Annual wobble (AW), as well as more accurate identification of the CW's phase jumps; For the $\Delta J_2$ record, we mainly focus on the interannual to interdecadal or even secular variations contained in it.

## 4.1 Application I: The fine structures of CW and AW in Polar motion

The Polar motion (PM) time series mainly consisted of two signals, the CW (with ~433day period) and AW (with 365day period). In light of the above-mentioned simulated test that the HSA can almost completely strip and restore the two input signals with time-varying amplitudes and periods, we can safely use the HSA to process the observed PM time series. The study of Polar motion structure contains two aspects: the precise extraction of CW, AW and the fine features of CW. On the one hand, the properties of refined PM periodic oscillation stripped from the observed polar motion are very important for studying their excitation mechanism and physical processes. For example, Smylie et al. (2015) applied the maximum entropy method (MEM) to reveal details of the effect of earthquakes on polar motion. On the other hand, the CW is regarded as the main eigenmodes of the Earth's rotation, and the investigation of its properties such as period, amplitude, and frequency variations are of equal importance for the understanding of the physical processes in the Earth (Malkin and Miller, 2010). Among these CW properties, a much interesting peculiarity is the phase jumps. It's Orlov (1944) who for the first time detected ~180° phase jump that occurred in the 1920s. Besides this most well-known CW phase jump, other significant phase jumps in the 1850s and 2000s have also been reported recently (Malkin and Miller, 2010). Some researchers have interpreted these CW phase jumps as anomalously occurring in temporary coincidence with geomagnetic jerks (e.g., An and Ding 2022), or Free Core Nutation (FCN) phase perturbations (Shirai et al., 2005). However, the CW phase's evaluation is a more complicated task than the CW extraction. For conventional filters and mathematical tools, it is a tough matter to separate the phase variations from the period variations. Fortunately, the additional function of the HSA algorithm supports estimating such immediate parameters.

Here we use the EOP C01 PM series during 1900–2020 with a 0.05-year sampling interval (http://www.iers.org) as the dataset, after applying a band-pass filter to it (with 0.2 and 2 copies as the cut-off frequencies), we denote the residual dataset as PM_R for further using. We refined CW and AW components from PM_R dataset by using HSA as an initial analysis. As shown in Fig. 8 (a), the AW variation has a significant trough in the 1990s when its amplitude reached the minimum value. From on, the AW amplitude begin to increase to around 100 mas and trended to be stabilized since the 2010s. These conclusions are the same as King and Agnew (1991)'s result. In addition, Fig. 8b shows the amplitudes of the CW currently are at a minimum level in history. The CW amplitude variations have been rapidly decreasing from 1995 to 2015, but a stable trend emerged after that. This result is different from some speculations that the reduced CW will be sustained to even diminish in 2022 yr (Wang et al., 2016). Another marked amplitude minimum in the 1920s has been found. As in the 1920s, the famous CW phase jump occurred accompanied by a trough in the CW amplitude (Shirai et al., 2005). For the explorations of both two CW jumps, the following HSA spectrum will further give a more detailed analysis.

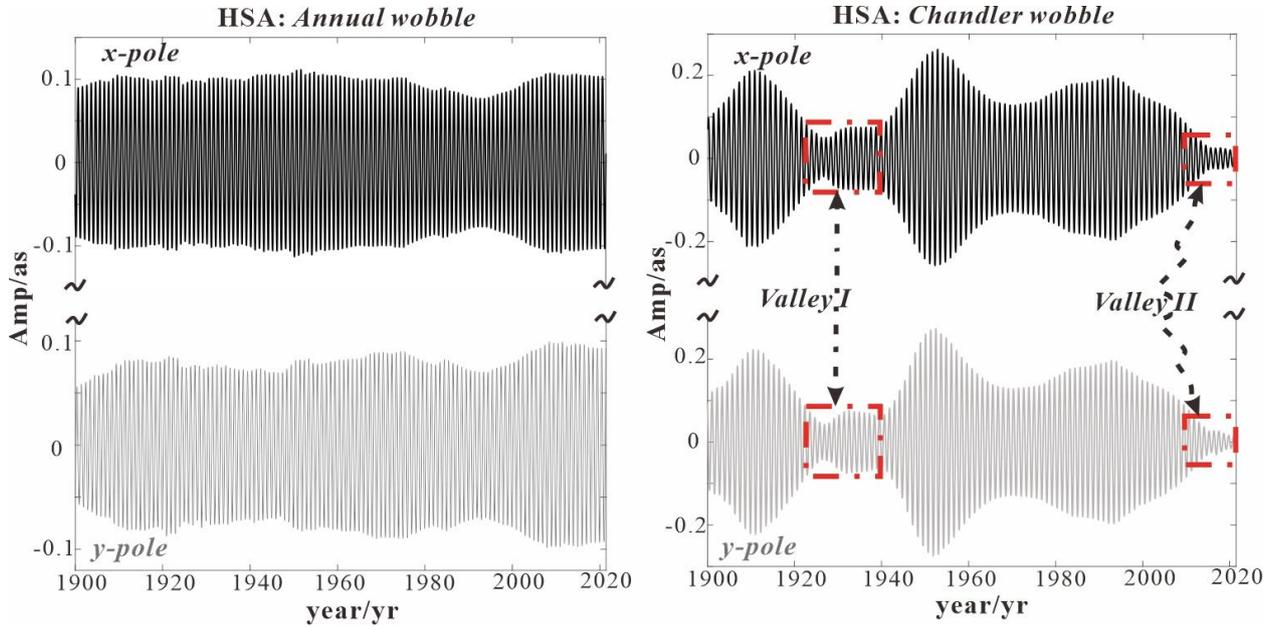

**Figure 8.** Reconstruction of the AW (a) and CW (b) using the HSA method

With the HSA Spectrum $\{A_i, f_i, \theta_i\}$ (see Fig. 9), the instantaneous phase, period and amplitude of Chandler wobbles will be revealed clearly. There are three impressively anomalous changes (denoted by gray areas in Fig. 9) that can be found in the HSA spectra. The first and most famous phase jumps occurred around the 1920s (Chao and Chung, 2012). From Fig. 9c, we can find that the large phase jump occurred around 1926yr, and the corresponding phase jumps are ~100° and ~150° for the *x*-pole and *y*-pole, respectively; meanwhile, it is clear that the amplitude and period of the CW also have local minima around the 1926 yr (see Figs. 9a-9b). Another two epochs with similar behaviors occurred around 1940 yr and 2015 yr. For the former, the subtle 1940yr phase jump is firstly observed more than ever and also synchronously manifested with a sharp decrease in amplitude and period (see the gray areas in Figs. 9a and 9b). As for the valley in 2015 yr, it remains more controversial (such as the occurrence discrepancies) caused of its location at the edges of the interval covered by the EOP C01 series (Malkin and Miller, 2010). These disputes are generally explained by different edge effects of the methods used. Based on the excellent reconstruction of HSA even at the ends of time series, we can firmly believe the existence of the phase jump in 2015 yr, which is also accompanied by deep minima of CW amplitude and period.

Using the excitation series, Chao and Chung (2012) have previously found it happened to be the opposite strength and direction of Chandler wobble and hampered the Chandler motion momentarily in the 1920s. Correspondingly, a new phase is generated, causing the phase jump in the Chandler wobble series. In this study, we find that there are three moments (1925, 1940 and 2015) when the amplitude and period of the CW simultaneously rapidly drop to a minimum, and all of them are accompanied by significant phase jumps. Our results, thereby, support his view of the non-unique Chandler phase jumps may as the consequence of the sharp decrease of Chandler intensity at the time, rather than the Earth system anomaly.

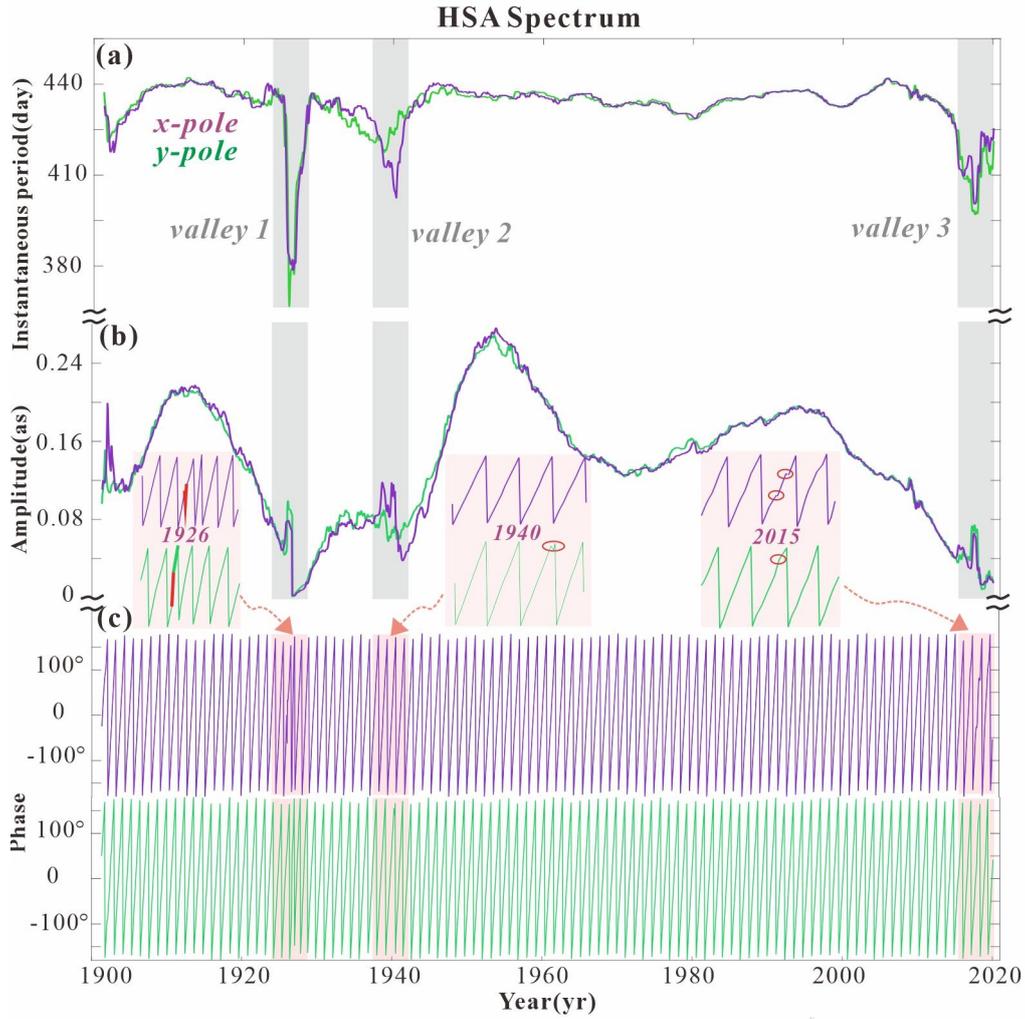

**Figure 9.** Instantaneous frequency (a), period (b), and phase (c) of CW in the presentation of the HSA spectrum

**4.2 Application II: The interdecadal periodicity of dynamic oblateness**

Since the satellite laser ranging (SLR) technique, Yoder et al. (1983) firstly associated the $\Delta J_2$ with the mass redistribution and gravity variation and reported a secular decrease of $\sim -3\times10^{-11}$/yr. With the accumulated SLR data, more long-term $\Delta J_2$ variations were concentrated by the present study, mainly aimed at the inter-annual or -decadal periodicity. In this application, we resorted to the HSA decompositions with the largest window ($L$=series length, equal to the linear HSA) to the $\Delta J_2$ low-frequency signal.

We employed the dynamical oblateness $\Delta J_2$ data produced from the Center for Space Research (http://download.csr.utexas.edu/pub/slr/degree_2/) in the availability for the 1976-2021 yr monthly time series (courtesy of M. K. Cheng; cf. Cheng et al., 2013). Because of that we only focus on the low-frequency oscillations, so we first filter the original time series. After applying a low-pass filter to the $\Delta J_2$ series (using 0.5cpy as the cut-off frequency), a residual time series $\Delta J_2\_R$ is obtained for the HSA analysis. For a much more complete fitting, we use up to 8 pairs of harmonics (order=16) for the $\Delta J_2\_R$ series and three pairs of damped harmonics (order=6) for the $\Delta J_2$ series secular approximation. In this case, double interesting periodicities emerge in the linear HSA spectrum (seeing the right panel of Fig. 10), and correspondingly their components are reconstructed in the left panel (marked in red line). Specifically for these peaks, the prominent 18.6yr signal is initially reconstructed from the $\Delta J_2$ residuals for the possible action of tidal effect. According to the IERS Convention 2010 (Petit & Luzum, 2010), the lunar orbit nodal precession excited the 18.61-year tide (solid tide of the elastic Earth as dominator) and is known to affect $\Delta J_2$. Fig. 10a hence gives the estimate of the solid tide effect. The well consistent implies the tide effect in elastic earth (IERS) that mainly accounts for the bulk of the observed $\Delta J_2$ 18.6 signal, but not completely with a slight phase difference of the 0.41yr (8.05°, referenced to epoch 2000.0). In Chao et al. (2020)'s study, the difference has been attributed to the other modified effects (e.g., mantle anelasticity, Eanes, 1995)

The HSA method also resolves another periodicity of yet-unclear origin at 10.5yr and the long-period signals responding to the recent secular variations. In secular scales, the existence of its peaks (a likely peak of 47.6yr or ~56yr by Yu et al., 2021, Ding and Chao, 2018) has been a controversial issue. Since 2005, the overall $\Delta J_2$ time series has significantly departed from the secular decrease to an upward swing, signifying the quadratic undulation. Because of it, Cheng et al.

(2013) have proposed that these interdecadal peaks may rather like be from the quadratic-term false mode. Due to the limitation of short-length observations, it is difficult for HSA to fully separate the ultra-low frequency signals, while the corresponding components will be manifested in a no-linear trend. Such a no-linear trend may be caused by the global sea-level rise and ice melting in the cryosphere (Cheng and Ries, 2018; Loomis et al., 2019) largely offsetting the $\Delta J_2$ downward trend. For the ~10.5yr signal, only one possible causality of the solar cycle was suggested after analyzing the sunspot variations (e.g., Chao et al., 2020). After using the HSA algorithm to process the sunspot series, however, the 10.7yr solar cycle has a relatively large phase deviation with the ~10.5yr $\Delta J_2$ signal (seeing Fig. 10), which implies that the yet-uncertainty of their causality. There are some other synchronous relationships with gravity anomalies seldom exposed. For example, a well-known "1998 anomaly" gravity field has been reported when $\Delta J_2$ reversed its decreasing trend for several years; but interestingly, we found that such $\Delta J_2$ peaks also appeared near 1982, 1990, and 1998 (see Fig. 10g, after de-quadratic). These local anomalies correspond to the integration of the wave humps of ~10.5yr, ~18.6yr and some interannual components (4~8 yr, marked in vertical arrows). In early studies, Chao et al. (2020) have found a significant interannual frequency dependence of $\Delta J_2$ with climatic oscillations (e.g., Antarctic and Arctic Oscillation); and Cheng and Tapley (2004) further speculated that some oscillation coupling (~ 6 yr, decadal variations) may lead to gravity anomaly. Here, for the first time, we observed the prominent multi-peaks of $\Delta J_2$ series (not only at 1998 yr); and such "anomalous" will frequently occur as they are mainly contributed by interannual-to-decadal periodicities.

Generally, direct identifications and restorations of those long-period signals are tough works, considering the limited number of data and low SNR; even for the 18.6yr tidal signal and 10.5yr signal in $\Delta J_2$ time series. Utilizing complex $z$-transform, the AR-z frequency spectrum implements a better identification in $\Delta J_2$ low-frequency points than the Fourier spectrum; and Chao et al. (2020) have attempted to implement a similar signal restoration with the AR-z prior information. In this study, we propose a powerful HSA method, which can be used for not only identification but direct restoration of the 18.6yr tidal signal and 10.5yr signal in the $\Delta J_2$ time series.

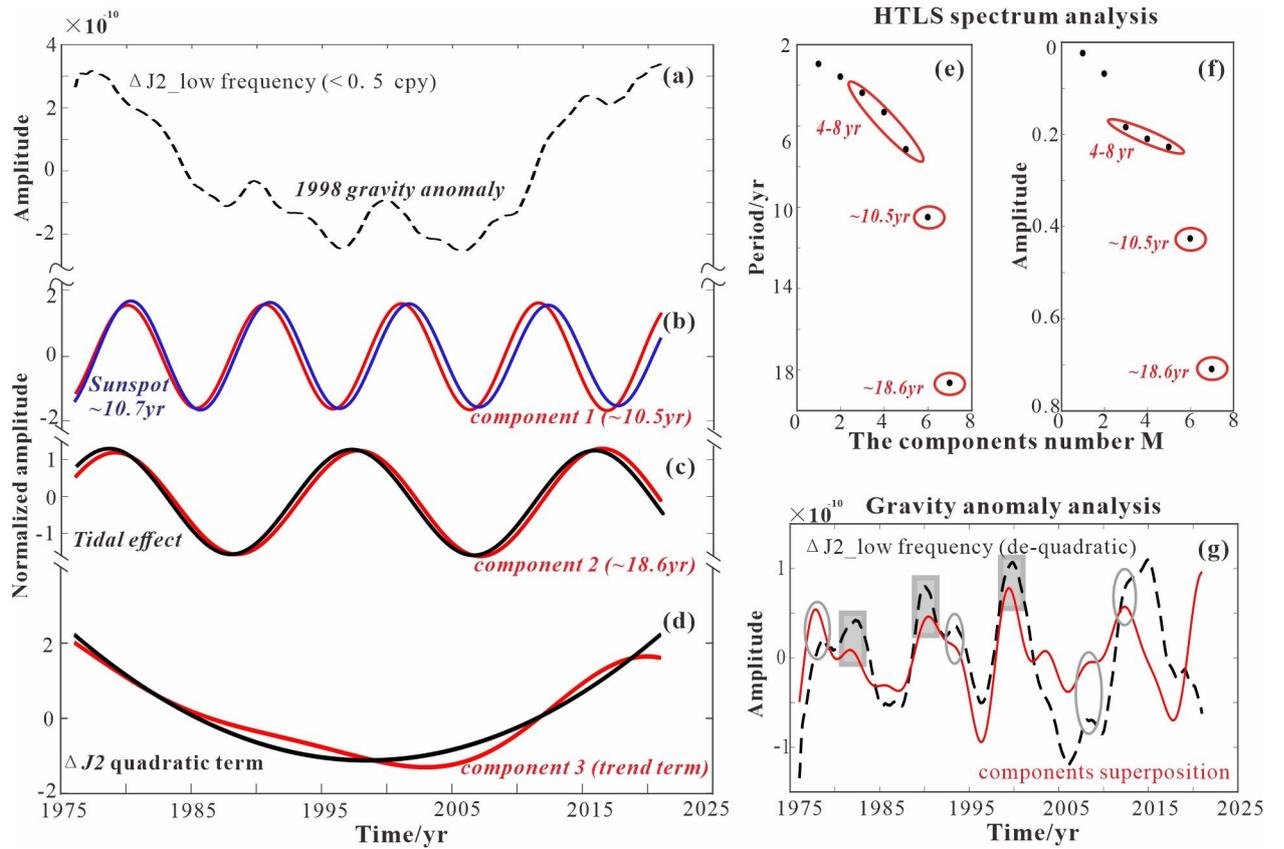

**Figure 10.** The HSA decomposition of periodicities from $\Delta J_2\_R$ series (a). The linear HSA spectrum records the 10.5 yr, 18.6 yr, and interannual signals (e-f); and our reconstructed results (marked in red curves) were compared with those by sunspot 10.5 yr signal (b), tide effect from theoretical elastic earth (IERS) (c) and quadratic term (d). (g) further shows the synchronous peak-to-peak in 1982, 1990, 1998 (marked gray square) and other less prominent stages in 1978, 1994, 2008, 2012 (marked gray ellipse) by the $\Delta J_2\_R$ series (removing quadratic) and a sum of 10.5 yr, 18.6 yr and interannual variations (4~8 yr).

# 5 Discussion and Conclusions

This paper develops a technique (HSA) for accurately retrieving and decomposing the non-stationary signals in geophysical observations. The essential work of HSA mainly involves the $z$-pole approach which dated back to the well-known Prony's relation 200 years ago. In his proposed complex harmonic functions, each $z$-poles corresponds to an exponential sinusoid and can be resolved by the AR model to polynomial $H(z)$. With an ill-conditioned matrix where the root number of $H(z)$ is usually large, unfortunately, the sensitivity to the AR model's coefficient perturbation will become much high. Here, we would like to offer a simple, analytic, and much more high-robustness method for $z$ poles solution by a combination of rotation invariant properties and a state-subspace of Hankel matrix arrangement. Compared with AR-$z$ poles, the Hankel-$z$ poles can mandatory ensure the poles inside the complex $z$-plane and effectively avoid exponential divergence happening.

To upgrade the decomposed performance in cases of gradual structure changes or transient conditions, we further resort to the technology of the window process. In previous studies, the idea has been successfully applied in the Short-Time Fourier Transform (STFT) and the Wavelet Transform (WT) (Walden and Cristan, 1998; Allen and Rabiner, 2015). However, the time and frequency resolutions of such methods depend strongly on the type and the length of the window (Bayram and Baraniuk, 2002). We break through the limitation by proposing an HTLS basis with a sliding window to track the transient harmonic variations. The HSA algorithm thereby is available to geoscientific observables with multiple time scales.

In the interannual and annual scales, the Annual wobbles and Chandler wobbles (with their jumps events) in polar motion are investigated as a case study. The most controversial topics focus on the CW properties especially recently, more research demonstrated the non-uniqueness of the CW phase jumps. With HSA Spectrum, the uncovered instant phases, periods, and amplitudes of the CW are conductive to deeply explore those questions. Here, we traced triple special epochs in Chandler wobbles, the 1920s, 1940s, and 2010s, all of them correspond to a phase jump; meanwhile, the amplitude and period of the CW simultaneously rapidly drop to a minimum. A prevailing assumption of the phase jump trigger is from FCN phase perturbations (e.g., Guinot, 1972; Gibert and Le, 2008), but it cannot account for the synchronous react of deep minima in amplitude and period. So what is the cause of similar behavior of the CW amplitude and period? and whether the phase jump and current deep minima of CW intense is related to the earth's internal structure? Our results demonstrated that these non-uniqueness CW jumps are worth further study.

In the interdecadal or aperiodic scale, we also did another analysis of the earth's oblateness $\Delta J_2$ variations. Usually, different used methods will lead to different sensitivity to the particular signals. Especially for the weak long-period $\Delta J_2$ signals, their identifications and reconstructions are rather hard to work because of the limited length of observations and low SNR. Our results revealed an obvious secular derived from HSA, and the other two non-climatic decadal oscillations. In these long-period signals, the $\Delta J_2$ variations at the ~18.6-year period related to the tidal effect (theoretical elastic earth, from IERS Convention 2010) are notably detected. Another potential relevance between the ~10.5yr signal and the solar cycle is also found (Chao et al., 2020). Though their correlation is early attributed to the modeling of solar radiation, the large phase deviation implies the uncertainty presently. Afterward, we discovered that some components (for interannual timescales of 4-8 yr, ~10.5 and 18.6 yr signals) can account for the well-known gravity field's "1998 anomaly", and also for some other 'anomaly' in 1982, 1990, and 2008 epochs.

In conclusion, these preliminary investigations in geophysical datasets of the Earth exhibit the HSA's potential as an effective tool to resolve the fine structures of non-stationary signals. No matter for the principal component separation, the weak harmonic detection, or the long-term trend reconstruction, our obtained impressive results suggest that the HSA can help to unlock more underlying important phenomena contained in the geophysical observations of the geo-field.

## Acknowledgments

The used EOP C01 polar motion series, can be obtained by downloading freely from : http://www.iers.org; and the $\Delta J_2$ series can be downloaded from: http://download.csr.utexas.edu/pub/slr/degree_2/. This study is supported by the National Natural Science Foundation of China (Grants: 41721003, 41974020, 41974022, and 41304003).